\documentclass[aps,prl,twocolumn,groupedaddress]{revtex4}

\usepackage{graphicx}
\usepackage{dcolumn}
\usepackage{bm}

\begin{document}

\title{Retrapping Current, Self-Heating, and Hysteretic Current-Voltage Curves \\in 
Ultra-Narrow Superconducting Aluminum Nanowires}

\author{Peng Li}
\author{Phillip M. Wu}
\author{Yuriy Bomze}
\author{Ivan V. Borzenets}
\author{Gleb Finkelstein}
 \author{A. M. Chang}%
 \email{yingshe@phy.duke.edu}
\affiliation{ Department of Physics, Duke University, Durham NC, 27708}

\date{\today}

\begin{abstract}
Hysteretic I-V (current-voltage) is studied in narrow Al 
nanowires.  The nanowires have a cross section as small as 50 
$\rm{nm^2}$.  We focus on the retapping current in a down-sweep of the 
current, at which a nanowire re-enters the superconducting state 
from a normal state.  The retrapping current is found to be 
significantly smaller than the switching current at which the 
nanowire switches into the normal state from a superconducting 
state during a current up-sweep.  For wires of different lengths, 
we analyze the heat removal due to various processes, including 
electronic and phonon processes.  For a short wires $1.5 \mu m$ 
in length, electronic thermal conduction is effective; for 
longer wires $10 \mu m$ in length, phonon conduction becomes 
important.  We demonstrate that the measured retrapping current 
as a function of temperature can be quantitatively accounted for 
by the self-heating occurring in the normal portions of the 
nanowires to better than 20 \% accuracy.  For the phonon 
processes, the extracted thermal conduction parameters support 
the notion of a reduced phase-space below 3-dimensions, 
consistent with the phonon thermal wavelength having 
exceeded the lateral dimensions at temperatures below $\sim 1.3 
K$.  Nevertheless, surprisingly the best fit was achieved with a 
functional form corresponding to 3-dimensional phonons, albeit 
requiring parameters far exceeding known values in the 
literature.

\end{abstract}

\pacs{Valid PACS appear here}
\maketitle

Understanding the dynamics of ultra narrow supeconducting (SC) 
nanowire wires is an active area of investigation 
\cite{Giordano.PRL,Nature,Lau.PRL,Highfield.B,Fabio.PRL,HysteresisIV,NaturePhysics,PengPRL,Goldbart,Little,LA,MH,QPStheory.Zaikin1,QPStheory.Zaikin2,sergei}.  
A significant focus is the so-
called 1-dimensional (1D) limit, delineated by the condition, 
$(w,h) < \xi$, where w is the width and h the height of the 
nanowire, and $\xi$ the superconducting coherence length.  
Investigations of the behavior under current-biasing 
not only elucidate the conditions and limitations for the 
current carrying capabilities, as well as the process of recovery back 
into the superconducting (SC) state after driven normal by an 
excessive current, but also potentially lay the foundation and 
pave the way for the development of novel devices, such as a 
current-Josephson effect devices \cite{CurrentJJ}, or qubits 
\cite{nanowire.qbit}.

In this work, we report on measurements carried out in ultra narrow Al nanowires with a cross section 
as small as 50 $\rm{nm^2}$.  
The three nanowires studied have a widths and heights ranging 
between 7 - 10 $\rm{nm}$, and lengths of 1.5 $\rm{\mu m}$ (wire S1) or 
10 $\rm{\mu m}$ (wires S2, S4).   These nanowires are exceedingly uniform 
in their cross section, as indicated by their ability to carry 
sizable current before being driven normal, where the current 
density is nearly identical to co-evaporated 2d films.  In a 
previous work, the behavior of the switching current $I_s$ 
during an up-sweep of the current was investigated \cite{PengPRL}.  
There, it was found that heat deposited by phase-slips--
transient temporal-spatial events during which the 
superconducting phase fluctuates and changes by $2\pi$ over a 
distance of order $\xi$, while the core region goes normal--leads to a 
thermal runaway, driving the entire nanowire into a normal state 
from the SC state.

\begin{figure}
\includegraphics[width=0.5\textwidth]{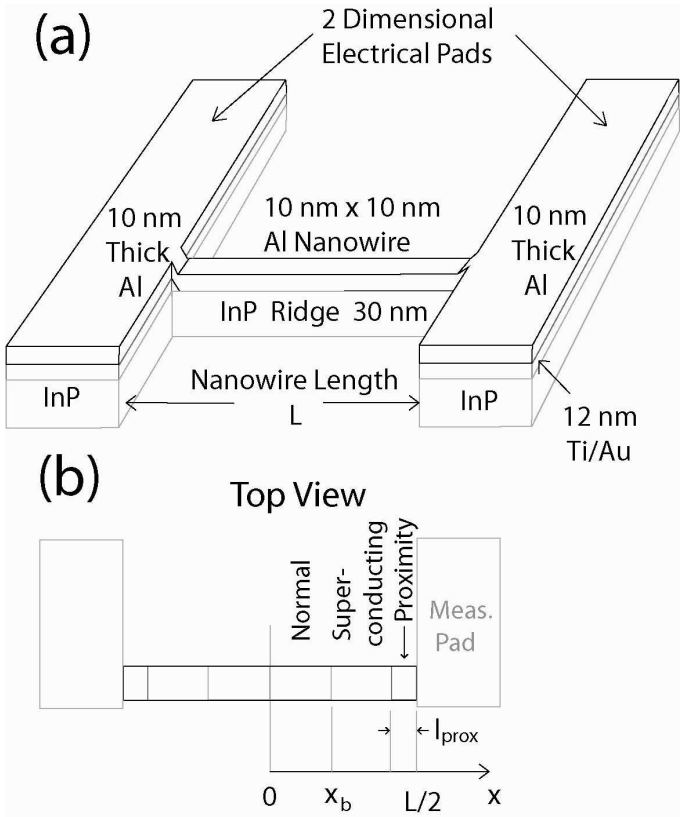}
\caption{\label{fig1} (a)  Schematic of the Al superconducting 
nanowire device on a narrow InP ridge template.  The ends of the 
nanowire are connected to large, electrical measurement pads.  The 
pads can either be in the superconducting state, or driven normal 
by a magnetic field.  (b)  Top view of the nanowire geometry, and 
the layout used in the heat flow model discussed in the text.}
\end{figure}

Here we focus on the down-sweep retrapping current.  The retrapping 
current $I_r$ is found to be 
significantly smaller than the up-sweep switching current $I_s$, 
and can be a much as a factor of 20 smaller.  
The history dependent current-voltage (IV) relation 
exemplified by the disparate behaviors in the up- and 
down-sweep is ubiquitous, despite the fact that based on the
criteria normally applied to SNS (superconductor-normal metal-
superconductor) bridges, the nanowires should be in the heavily 
over-damped regime in its dynamics \cite{HysteresisIV,NaturePhysics,Hysteresis.SNS}.   
In $\rm{MoGe}$ nanowires of widths $\sim 10 nm$, 
Tinkham {\it et al.}\cite{HysteresisIV} performed a heat flow analysis, 
and ascribed the retrapping behavior to self-heating.  Our work bears 
similarity to that work, but our SC nanowires are in a different 
regime, where  $k_F l \sim 60 \gg 1$, rather than being close to 
1 in their case.  Here $k_F$ is the Fermi wave-number, 
and $l$ is the mean-free-path.  Moreover, their nanowires were 
suspended freely, while ours are deposited onto a narrow, 
8 nm-wide InP ridge (Fig. ~\ref{fig1}(a)), and are thus in 
thermal contact with an underlying substrate.  Furthermore, our 
analysis differs from theirs in the form of 
the heat flow equations.  Based on our analysis, we rule out under-damping 
as the cause of the hysteresis, in agreement with recent results in 
submicron SNS bridges \cite{Hysteresis.SNS}.  

\begin{figure}
\includegraphics[width=0.5\textwidth]{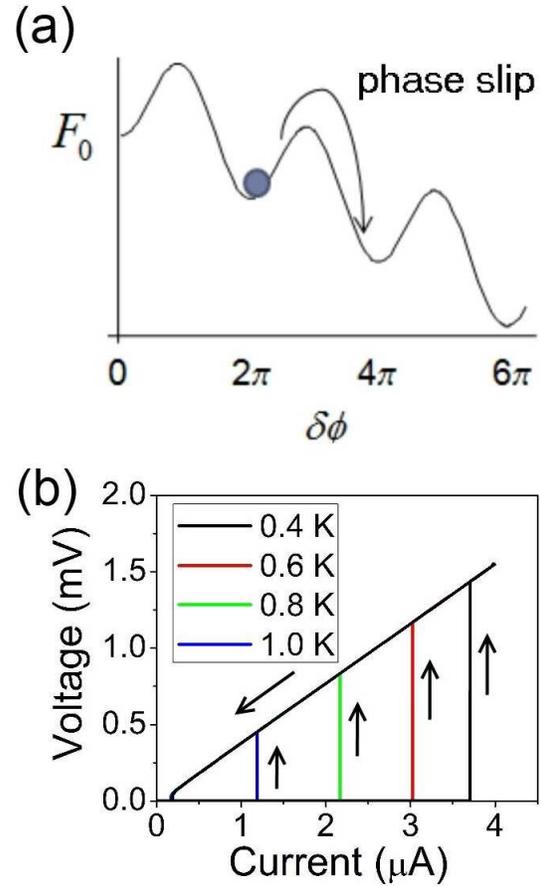}
\caption{\label{fig2} (a) Tilted-washboard free energy landscape for 
a Josephson junction under current bias.  A phase slip occurring between 
adjacent minima is indicated.  A similar scenario occurs in a 1D 
superconducting nanowire.  (b) Hysteretic IV curves for nanowire S2 at several 
temperatures.  Arrows indicate direction of current sweeps.  The upsweep 
switching current $I_s$ is significantly larger than the down-sweep 
retrapping current $I_r$. }
\end{figure}

To lay the framework for understanding the behavior of 
nanowires, the Josephson junction can serve as a starting point.  
There, the free energy landscape under current bias is described 
by the tilted washboard potential, shown in Fig. ~\ref{fig2}(a) 
\cite {TinkhamBook}.  This same scenario is also applicable to 
1D SC nanowires \cite{LA,MH}.  Josephson junctions are classified 
within a Resistively and Capacitively Shunted 
Junction (RCSJ) model as either under- or over-damped, 
depending on whether the quality factor, 
$Q=\sqrt{2eI_cC/\hbar}R$, is greater or less than 1.  Here, 
$I_c$ is the critical Josephson current, C is the junction 
capacitance, and R the junction normal state resistance.  When  
under-damped Josephson junction is driven over the free-energy 
barrier out of its meta-stable minimum, the SC phase keeps running 
downhill as there is insufficient damping to retrap the phase in 
a lower energy local minimum.  A consequence is that a 
hysteretic current-voltage (IV) relation, where the up-sweep and 
down-sweep branches do not overlap.  In contrast, in an over-damped junction, 
the phase moves diffusively between adjacent minima, and hysteresis is 
often not present \cite{MQT.Clarke,Collapse1,Collapse2}.  

The estimated $Q$ for our nanowiresis in the range of $\sim 
0.01$, far below unity, and the nanowires are ostensibly 
in the severely over-damped limit. This estimate is relevant 
when the nanowire device is in the S-NW-S configuration, where 
S refers to each of the two large metallic measurement pads 
when in the SC state, and NW denotes the nanowire.  It also 
provides a reasonable estimate in the N-NW-N configuration,  
when the pads are driven normal, but the ambient temperature 
is below the nanowire SC transition temperature $T_c$.  In 
this case, the nanowire itself breaks up into alternating 
SC and normal segments, whether during an upsweep or a downsweep 
of the current.  In the former case, the nanowire is overall in 
the SC state, but during a phase-slip, the phase-slip normal 
core acts as the normal region.  In the latter, the central 
portion of the nanowire is normal due to heating, while the 
regions closer to the pads are in the SC state (Fig. 1(b)).

Nevertheless, despite the over-damping, hysteretic IV curves are 
ubiquitous, as can be seen for wire S2 in Fig. ~\ref{fig2}(a).  In 
fact, the ratio of $I_r$ to $I_s$ can be as small as $\sim 1/20$.  
For example, in nanowire S2 at $T \sim 0.3 K$, $I_r \sim 0.19 \mu A$, 
while $I_s \sim 4 \mu A$.  These
observations motivated us to investigate the retrapping current 
systematically, as a function of the temperature and wire length, and to perform a detailed heat 
analysis to establish self-heating as a cause of the substantially reduced $I_r$ below the value of the upsweep 
$I_s$.  

Our devices were fabricated using a template method.  The 
template is a narrow, $\rm{8 nm}$ wide InP ridge, formed by 
differential etching on the cleaved (110) crystallographic plane 
of a molecular-beam-epitaxy (MBE) grown InGaAs-InP crystal, where   
the growth direction is (001).  The geometry of our devices is 
depicted in Fig. ~\ref{fig1}(a).  The details of the fabrication 
procedure is described in a previous work \cite{Fabio.APL}.  
The nanowire resides on the narrow InP ridge and is thus 
thermally connected to the large semiconductor substrate through 
the narrow ridge.  The nanowire is electrically connected 
to large metallic measurement pads on its ends.  Therefore, for 
heat removal, thermal conduction both in the lateral direction 
along the nanowire, and vertically through the InP ridge via 
phonon processes must be considered.  The IV measurements were 
carried out in a extremely carefully shielded apparatus to 
minimize unwanted environmental interference, such as external 
noise (e.g. from nearby radio stations) conducted down the 
electrical cables, or Johnson-Nyquist noise from resistors 
within the electrical measurement circuitry.  In particular, 
Thermax cables with the ability to remove high frequency 
noise is employed where possible, as well as low-temperature RF 
filters.  The devices are also enclosed in metal cans with all 
openings plugged with conductive tape or metal mesh.

\begin{figure}
\includegraphics[width=0.5\textwidth]{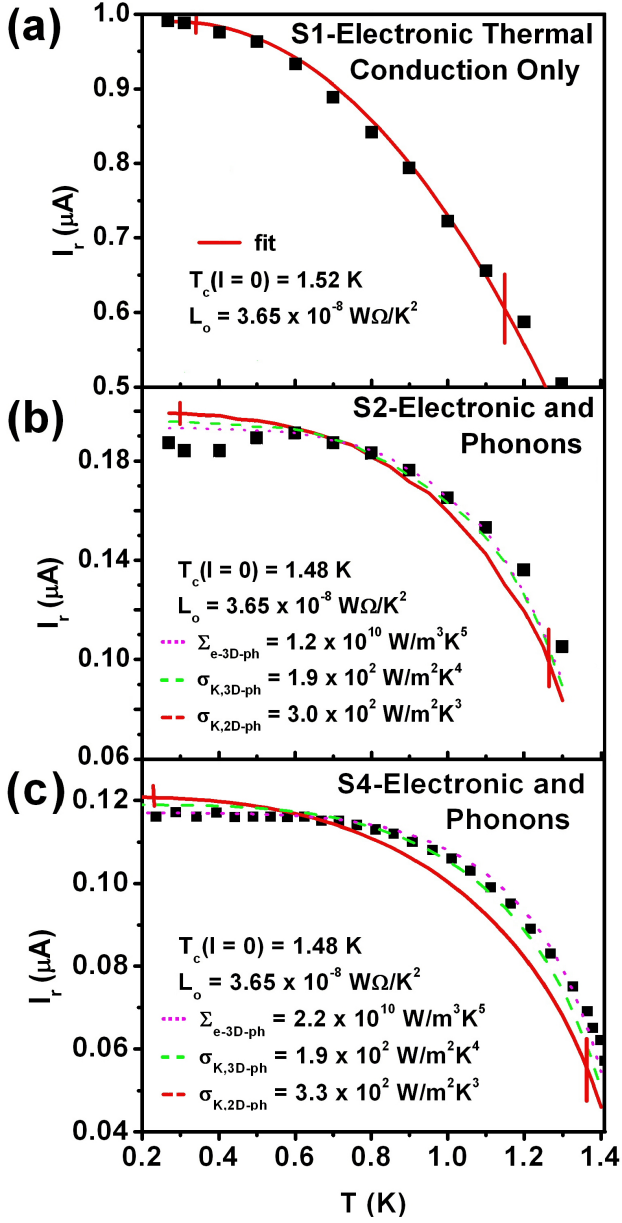}
\caption{\label{fig3} Solid squares--Data for the retrapping current 
$I_r$as a function of temperature for nanowires (a) S1 ($\rm{10 nm \times 
10 nm \times 1.5 \mu m}$), (b) S2 ($\rm{9.3 nm \times 9.3 nm \times 10 \mu m}$), 
and (c) S4 ($\rm{7 nm \times 7 nm \times 10 \mu m}$).  
For S1 in (a), 
the solid curve is a fit based on the heat flow model describe 
in the main text, 
where only the electronic thermal conduction is considered.  The 
Wiedermann-Franz constant deduced from the fitting is 
$L_o = 3.65 \pm 0.15 \times 10^{-8} W \Omega /K^2$.  
Vertical bars indicate certainty of the fitting curve, due to a 
combination of the uncertain in the parameters, including uncertain in 
the superconducting transition temperature $T_c(I)$ of $\pm 2 \%$.  
For S2 and S4 in it is necessary to include both electronic 
and phonon conduction.  Various rate limiting phonon processes were 
considered, including electron phonon relaxation and Kapitza 
boundary resistance phonon conduction (see text).
The uncertain of the phonon parameters is $\pm 10 \%$.  
%
}
\end{figure}

In Fig. ~\ref{fig3}(a)-(c), we present the measured $I_r$ for 
the three nanowires S1 ($\rm{10 nm \times 10 nm \times 1.5 \mu m}$), S2 
($\rm{9.3 nm \times 9.3 nm \times 10 \mu m}$), 
and S4 ($\rm{7 nm \times 7 nm \times 10 \mu m}$), respectively, 
as a function of temperature.  Immediately apparent is 
the substantial difference in the magnitude of the retrapping 
current $I_r$ for the short S1, when compared to the longer 
S2 and S4.  For S1, $I_r = 0.99 \mu A$ at $T = 0.3 K$, while 
it is $0.19 \mu A$ and $0.117 \mu A$, respectively for S2 and 
S4.  Thus, the value of $I_r$ in S1 is roughly 5 - 8 times 
that in S2 or S4.  The results for S2 and S4 do not depend 
on whether measurements were carried out in the S-NW-S or 
the N-NW-N configuration.  On the other hand, for S1, due to 
the larger current level and associated excessive heating, 
it was necessary to drive the measurement pads into a normal state, 
into the N-NW-N 
configuration.  By driving these pads into the normal state 
using a small magnetic field $B = 0.1 T$, the now normal pads 
can act as good thermal anchors, keeping the temperatures 
of the ends at the ambient temperature $T_o$.  For the 
longer S2 and S4, the smaller current level means that a much  
lower amount of heat needs to be carried out through wire ends; 
thus improved thermal anchoring was not needed.  For S1, on 
the other hand, keeping the pads SC (S-NW-S configuration) 
reduced $I_r$ to $\sim 0.55 \mu A$ from $\sim 1 \mu A$ as 
heat removal becomes more difficult due to poor thermal 
conduction capability of the SC pads.  At the same 
time, instabilities arise in the temperature profile along 
the nanowire, leading to very noisy data below $0.8 K$ 
with $I_r$ fluctuating as much as $0.1 \mu A$ between adjacent 
data points.

The configuration with {\it normal} electrical pads is relevant 
for our data in Fig. 3.  {\it In the analysis which follows, we will focus on 
this configuration.}   
The retrapping process returns a nanowire into the SC state during a down-sweep 
of the current, $I$, at an ambient lattice temperature, $T_o$, below the 
zero current critical temperature, $T_c(I=0)$.  At large $I$, most of the nanowire remains 
normal due to self-heating, which raises the local temperature above the switching temperature 
at that current, $T_s(I)$.  

During a current downsweep, the temperature is position dependent 
along the nanowire.  In the N-NW-N configuration, each of its 
ends is connected to a large, normal metal 
electrical measurement pads, anchored at $T_o$.  
Because $T_o < T_c(I=0)$, and is in fact below $T_s(I=I_r)$, 
the end regions are in the SC state, aside from 
a short proximity region, $l_{prox} \sim 100 nm$ in length, 
immediately adjacent to each normal pad 
(Figs. ~\ref{fig1}(b)).  Here, the $T_s(I)$ is the upsweep 
switching temperature at current $I$, On the other hand, as long 
as $I$ exceeded $I_r$, the center of the 
nanowire is above $T_s(I)$ and is thus in the normal state.  
When $I$ is reduced down to $I_r$, a blockage preventing the 
central region to be cooled is suddenly removed, allowing the 
cooling to propagate all the way to the center.  This blockage removal is described 
in detail in what follows.

The upsweep switching temperature for a given current $T_s(I)$ 
expresses the same relation as the switch current as a function 
of temperature $I_s(T)$, but viewed in reverse.  For our nanowires, 
these were reported in Ref. 8 \cite{PengPRL}.  
It is worthwhile to point out 
that the value of $I_s$ at a given temperature is not unique, 
but depends on the upsweep ramp rate of the current.  Conversely, 
$T_s(I)$ is also dependent on the ramp rate.  
This is due to the fact that the switching is caused by 
phase-slip events \cite{Goldbart,NaturePhysics,PengPRL}, and thus the relative 
rates of the phase-slip generation to current upsweep matters.  
The slower the upsweep current ramp rate, the longer waiting time 
is available for phase-slip 
events to take place within a given interval in current.  This 
increases the probability for switching within that interval and 
reduces the magnitude of the current at which switching takes 
place. But because the rate of phase-slip generation is 
exponentially dependent on the current, the dependence of $I_s$ on 
ramp rate is weak, and is approximately logarithmic.

To estimate $I_r$, it is necessary to determine the position dependent temperature, 
$T(x)$.  Both the electronic and phonon thermal conduction mechanisms need to be considered.  
The temperature range of interest is low compared to the lattice DeBeye temperature 
$\Theta_D \sim 300 K$ and the temperature dependence of the phonon thermal conductivity takes a power-law form, 
reflecting the phonon density of states.  At these low temperatures, $0.2 K < T < 1.3 K$, 
the value of the phonon conductivity is considerably smaller than the electronic thermal 
conductivity.  The only exception is in the SC regions when $ T(x) < 0.35 K$.  Thus, 
in the absence of a 
very large temperature rise phonons can only carry away a relatively small amount of heat.  
Whereas electronic conduction requires the heat to exit the ends of the nanowire, the phonon 
conduction goes through the short InP template ridge (in height), on which the 
nanowire resides.  The nanowire length is microns $\rm{\mu ms}$ while the ridge is only $\rm{30 nm}$ in height.  
If the wire is long, the electronic mechanism will become much less effective, and phonon 
conduction must be included as well.  

For the short wire S1, electronic conduction overwhelmingly dominates.  For the long wire S2, 
both electronic and phonon thermal conduction must take place side by side.  Phonon conduction 
through the InP ridge takes place via several steps:  (a) electron-phonon energy relaxation within 
the aluminum nanowire, (b) conduction through the aluminum-InP boundary, and (c) conduction through 
the 8 nm wide, 30 nm tall InP ridge.  Below the ridge, the energy is dissipated in the very highly 
conductive GaAs bulk material.  Thus the base of the InP ridge can safely be assumed to be held 
at the ambient temperature, $T_o$.  

One additional mechanism of heat removal takes place through the liquid $He^3$, which surrounded 
the nanowires in the set of measurement on S1 and S2.  However, this channel appears less important.  
Additional data for wire S4 obtained in the dilution refrigerator, in which the samples are in vacuum 
and thus there is no liquid surrounding the nanowire, yielded a retrapping current 
which can be accounted for in a similar manner as S2, using electronic and InP ridge 
thermal conduction only.  Note both wires S2 and S4 are $10 \mu m$ in length.

We divide a nanowire into two symmetric halves of length $L/2$ each, where $L$ is the total wire 
length, and consider the right half, where $ 0 \le x \le L/2 $, and $T(x=L/2) = T_o$, as shown in Fig. ~\ref{fig1}(b).  
When slowly down-sweeping the current $I$, we assume that the 
nanowire is in the SC state at a position $x$, if $T(x) < T_s(I)$, but is in a normal state if 
$T(x) > T_s(I)$.  

We begin by considering the short wire S1 ($1.5 \mu \rm{m}$ in length) and only include electronic conduction.  
Phonon conduction alone will remove $\sim 3\%$ of the heat generated by heating, and will be 
neglected.  The diffusion equations must account for three regions: 
(a) the central region for $0 \le x \le x_b$, which is normal for $I > I_r$ and has a resistance 
per unit length of $R_N/L \sim 0.33 -0.82 k\Omega/\mu m^2$ \cite{PengPRL} 
and thus self-heats; (b) the SC segment for 
$x_b \le x \le L/2 - l_{prox}$, which nearly does not self heat, but must conduct the heat generated 
by the central normal segment, and (c) the short proximity region adjacent to the normal metal pads 
for $ L/2 - l_{prox} \le x \le L/2$ which is approximated as a normal region.  More precisely, 
the SC region close to the SC-normal boundary $x_b$ heats slightly 
due to occasional phase slips as its temperature is just below $T_s(I) \ge 1.1 K$; the proximity 
region generates self-heating, and in addition must conduct through it the heat of the central 
normal region as well.  For the normal metal regions, the steady state heat diffusion equation 
is given by:
\begin{equation}
\frac {I^2R_N}{L} = - \frac {d}{dx} (\kappa_N A_{nw} \frac {dT}{dx}), 
\end{equation}
where $\kappa_N = L_oLT/(R_N A_{nw})$ is the electronic Wiedemann-Franz electronic thermal conductivity, 
$L_o$ is Lorenz number determined from fitting, and $A_{nw} = w \times h$ the nanowire cross sectional area.  
In the SC region, it is replaced by:
\begin{equation}
0 \approx - \frac {d}{dx} (\kappa_{SC} A_{nw} \frac {dT}{dx}). 
\end{equation}
The equations in the three regions are supplemented by boundary conditions at the 
junctions.  The junction between the center normal region and the SC region takes place   
at $x = x_b$, and between the SC and short proximity regions, at $x = L/2 - l_{prox}$.  
The temperature is continuous across each junction, and the heat flow 
is identical immediately to the left and right.  Lastly, we have $T(x = L/2) = T_o$.  Note 
that the forms of these equations 
differ from those used in Ref. 6.  There, the variation 
of the thermal conductivity $\kappa_N$ or $\kappa_{SC}$ with 
position, through their dependence on temperature T, was 
not accounted for \cite{HysteresisIV}.

\begin{figure}
\includegraphics[width=0.5\textwidth]{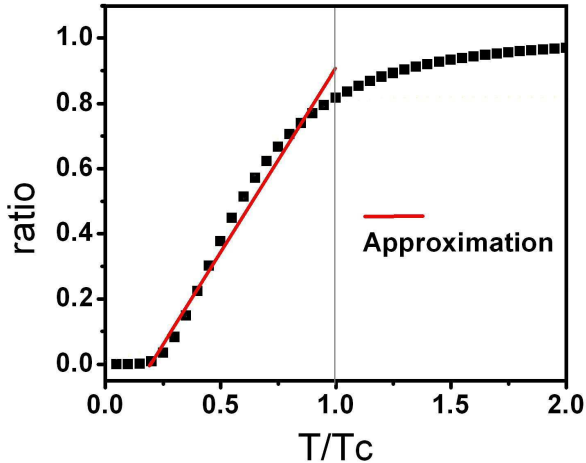}
\caption{\label{fig4} Plot of the thermal heat capacity ratio, $\kappa_{SC}/\kappa_N$, 
between the normal and superconducting states, as a function of reduced temperature, 
$T/T_c$.  The line indicates the approximation used in the calculations.}
\end{figure}

At a given temperature below $T_c(I)$, the electronic 
thermal conductivity of the SC segment at a position {x}, with a temperature $T(x)$, is 
related to the corresponding normal metal W-F thermal conductivity at that temperature, by the 
ratio $r \equiv \kappa_{Sc}/\kappa_N$:\cite{Hysteresis.SNS}
\begin{equation}
r = \frac{3}{\pi^2} \int^{\infty}_{1.76T_c(I)/T(x)} \frac{x^2}{cosh^2(x/2)} dx .
\end{equation}
This integral evaluated numerically is shown in 
Fig. ~\ref{fig4}.  
For the purpose of determining the position dependent temperature at a given $I$, this ratio 
is approximated by a linear form as indicated in the figure for $0.2 < T(x)/T_c(I) < 1$:
\begin{equation}
r \approx 1.125  [T(x)/T_c(I) - 0.2] .
\end{equation}

\begin{figure}
\includegraphics[width=0.5\textwidth]{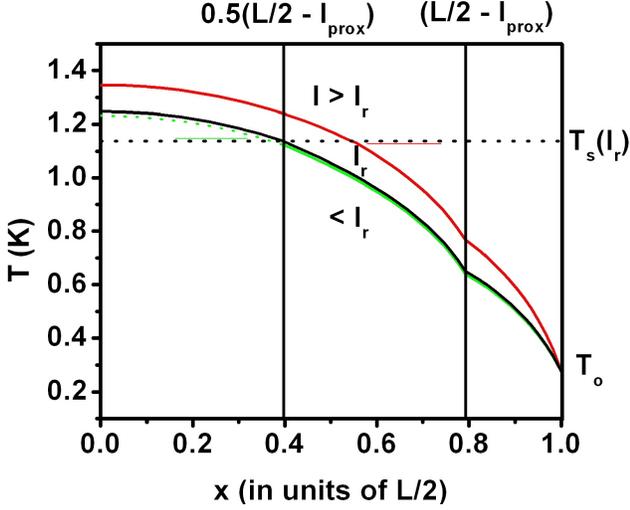}
\caption{\label{fig5} 
The temperature profile versus position x at ambient temperature $T_o$, with a 
corresponding retrap current $I_r$.  Three different current levels are 
depicted: $I > I_r$, $I = I_r$, and $I < I_r$.  The critical value of the 
SC-Normal boundary, is given approximately by $x_c = \rm{0.5(L/2 - l_{prox})}$.  
The proximity region lies to the right of $x = \rm{(L/2 - l_{prox})}$, ending at 
the electrical pad at L/2.  The SC segment lies immediately to the left.  For $I \ge I_r$, 
the position of the SC-Normal 
boundary, $x_b$, is given by the intersection of T(x) and the corresponding switching 
current $T_s(I)$.  The horizontal dotted line depicts $T_s(I_r)$, while the short 
light solid lines correspond to $I < I_r$ and $I > I_r$ with 
$[T_s(I > I_r)] < [T_s(I_r)] < [T_s(I < I_r)]$.  For $I > I_r$, $x_b > x_c$.  For 
$I = I_r$, $x_b = x_c$.  For $I < I_r$, indicated by the 
light solid curve joined onto the light dotted curve, the dotted "normal" 
region is unstable; the region immediately to the left of $x_c$ actually falls 
below $T_s(I < I_r)$ and will go superconducting, leading to a propagation of 
the SC-Normal boundary toward the center at $x = 0$.
}
\end{figure}

We next describe how the cooling blockage can be removed, and determine 
the condition for this to occur.  To do so, we consider the SC region and fix 
the current at $I$.  The SC-normal boundary occurs at $x_b$, which is determined by 
equating the temperature at $x_b$, $T(x_b)$, to the switching temperature at 
that current $T_s(I)$.  Momentarily treating $x_b$ as a variable, $T(x_b)$ attains 
its maximum value at a critical value $x_b = x_c$; in the simplest approximation 
$x_c$ is independent of $I$ as will be seen below.  For a large current $I$, 
the normal region is large and the actual $x_b$ exceeds $x_c$ (Fig. ~\ref{fig5}).  
As $I$ is reduced 
down to $I_r$, the normal region shrinks and $x_b$ becomes equal to $x_c$.  Here,  
$T(x_b = x_c)$ takes on the maximum possible value at $I_r$, since 
$x_b = x_c$, and is equal to 
$T_s(I_r)$.  Further reducing $I$ to just below $I_r$, $T_s(I < I_r)$ will 
slightly increase from $T_s(I_r)$, while at every $x$, $T(x)$ will slightly decrease 
due to reduced heating.  The decreased maximum temperature at $x_c$, $T(x = x_c)$, 
can no longer reach the increased $T_s(I < I_r)$.  The boundary will become unstable, 
and will propagate toward the center at $x = 0$.  Starting from the inital boundary 
at $x_b = x_c$, more and more of the normal region will fall below $T_s(I)$ and  
become superconducting, as the shrinking normal region generates less and less 
heat, until the entire wire is cooled.  The three cases are 
depicted in Fig. ~\ref{fig5}.

The value for $x_c$ can be deduced from the steady-state heat diffusion equation in the SC 
state, Eq. 2.  Neglecting the short proximity region adjacent  
to the pad, $x_c = L/4$ ($0.5(L/2)$); accounting for the proximity region of length 
$l_{prox}$ modifies this to $x_c \approx (L/2 - l_{prox})/2$.  
For illustrative purposes, let us determine $x_c$ in the absence of the proximity region.  
Focusing on the SC region at its border with the normal segment, 
$x = x_b$, twice integrating the diffusion equation and matching the boundary conditions 
yields for the left-hand-side (LHS):
\begin{equation}
LHS = (I^2R_N \frac {x}{L}) (L/2 - x) ,
\end{equation}
which is maximal for $x = L/4$ for fixed $I$.  At $x = x_b$ the factor 
$I^2R_N \frac {x_b}{L}$ represents the heating power generated by the normal region where  
$ 0 \le x \le x_b$.  Equating the LHS to a twice-integrated right-hand-side 
and solving for $T(x_b)$ thus yields the highest temperature at the normal-SC border 
when $x_b = x_c = L/4$, where the LHS is maximal.

\begin{figure}
\includegraphics[width=0.5\textwidth]{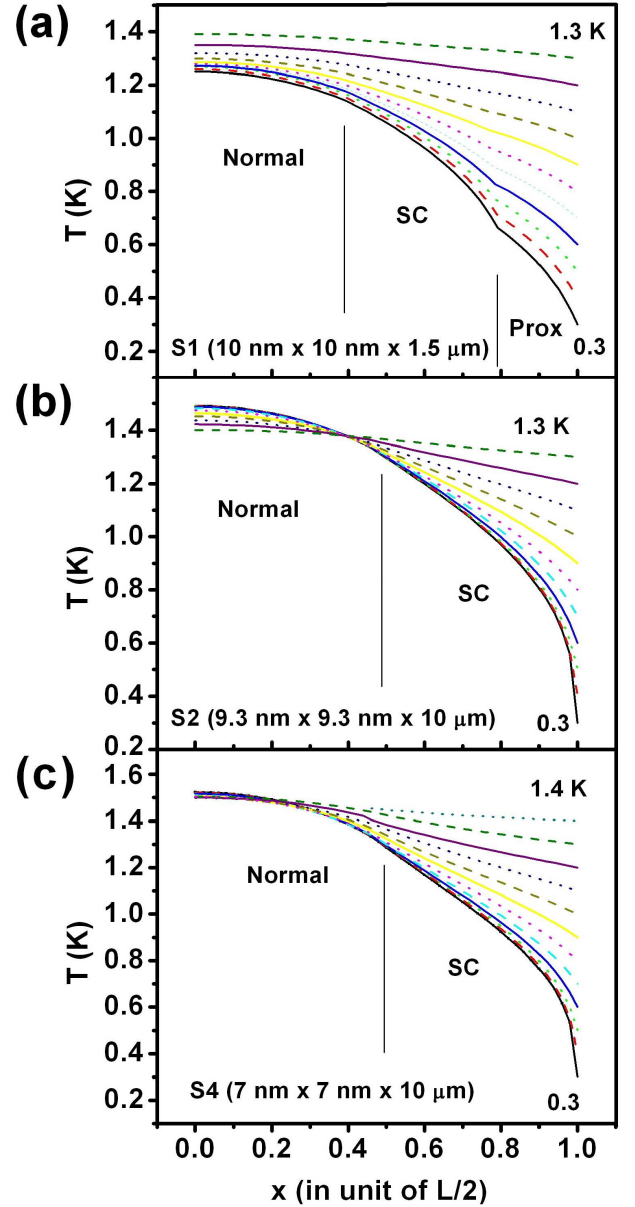}
\caption{\label{fig6} Temperature profile versus position x for different 
ambient temperatures.  The temperature is stepped by 0.1 K between 
successive curves--(a) nanowire S1, (b) nanowire S2, and (c) nanowire S4.  
The kink at $\rm{x = 0.79 (L/2)}$ in (a) is an artifact of modeling the proximity 
region as a normal metal, with a sharp boundary with the SC region to its 
left.  
}
\end{figure}

Using the approximate form of the thermal conductivity ratio $r$ between the SC and normal 
states given by Eq. 4, the diffusion equation can readily be solved analytically.  The solution yielded the position 
dependent temperature profile shown in Fig. ~\ref{fig6}(a).  
The kink at the SC-proximity boundary is an artifact of 
our model, where the proximity region is approximated as a 
normal region.  A more accurate model would require solving 
the Usadel equation, which is expected to yield a rounding 
and smoothing of the kink.  See Ref. 24 
for an example of the rounded voltage profile in the proximity 
region.  The fit to the $I_r$ as a function of temperature $T$ 
is presented in Fig. ~\ref{fig3}(a), using a Lorenz number $L_o = 3.65 \times 10^{-8} W\Omega/K^2$, somewhat 
higher that the theoretical value of $2.45 \times 10^{-8} W\Omega/K^2$.  
Viewed in another way, {\it forcing $L_o$ to take the theoretical value}, our 
model would predict a low temperature $I_r \sim 0.8 \mu A$, rather than 
the $1 \mu A$ we observed.  Reconciling 
this discrepancy may require the development of more sophisticated analysis using 
the Usadel equation, while incorporating heating and a position dependent temperature 
at the same time.  Despite the discrepancies, the overall 
behavior and magnitude (within 20 \% accuracy) are captured in 
our simplified model.  

For the $10 \mu m$ wires S2 and S4, electronic thermal conduction alone is not sufficient to support 
the measured $I_r$, despite its smaller value (by a factor of 5 - 8 at $0.3 K$).  It is 
necessary to include phonon conduction.  We 
assume a power-law temperature dependence in the heat exchange rate.  At every position, 
the heat-removal linear power density is set proportional to $T^{\gamma} - T_o^{\gamma}$, 
where $\gamma$ may be varied to reflect the limiting process in the phonon conduction discussed 
above.  The steady-state heat 
diffusion equation takes the forms:
\begin{equation}
\frac {I^2R_N}{L} - \beta(T^{\gamma} - T_o^{\gamma})= - \frac {d}{dx} (\kappa_N A_{nw} \frac {dT}{dx}), 
\end{equation}
and 
\begin{equation}
- \beta(T^{\gamma} - T_o^{\gamma}) \approx - \frac {d}{dx} (\kappa_{SC} A_{nw} \frac {dT}{dx}), 
\end{equation}
for the normal and SC segments, respectively.  For the longer wires S2 and S4, the smaller 
$I_r$ values allow the short proximity region adjacent to the normal-metal pads to be 
neglected.

This form for the heat-removal power-density-per-unit-length assumes that phonon conduction along 
the wire direction is weak--a reasonable assumption since the thermal conductivity is small, 
and since the wire is much longer than the height of the InP ridge (i.e. $\rm{10~\mu m \gg 30~nm}$).  
It also assumes that the environment to which the energy is dissipated is well-anchored 
at the ambient temperature.  For instance, if the electron-phonon relaxation is the 
limiting step, it is reasonable to assume that the combination of Kapitza and InP ridge phonon 
conduction is sufficiently large that the phonon temperature within the nanowire is maintained 
at $T_o$.  On the other hand, if Kapitza boundary resistance limits the rate of heat removal 
via the phonons, then it is reasonable to assume the phonons within the InP ridge 
are at $T_o$.

Several choices are possible for $\gamma$.  For electron-phonon energy relaxation via 
3-dimensional (3D) phonons, $\gamma = 5$, while for Kapitza boundary resistance, $\gamma = 4 $.  
However, the small lateral dimensions (width and/or height) in either the 
nanowire or InP ridge should render the phonons reduced below 3D, down to 1D and 2D, respectively.  
This occurs because the thermal phonon wavelength $\lambda_{ph}/2$, exceeds the lateral dimension(s) 
for $ T \le 1.3 K$ (more precisely, $\lambda_{ph}/2 > (w,h)$).  
Note that the width of the InP ridge is $w_{InP} = 8 \rm{nm}$ while 
the 
width/height of the nanowire $w$ or $h \sim 10 \rm{nm}$.  By examining the fitted coefficient in 
front of the power-law term, one seeks to exclude various possibilities.  This coefficient 
$\beta$ naturally depends on the limiting mechanism.  For example, in the case of electron-3D phonon relaxation, 
$\beta = \Sigma_{e-3D-ph} A_{nw} $, where $\Sigma_{e-3D-ph}$ is the 3D energy relaxation rate constant and 
$A_{nw}$ the nanowire cross sectional area.  For Kaptiza boundary resistance limited thermal 
conduction with 3D InP phonons, $\beta = \sigma_{K,3D-ph} w_{InP}$, where $\sigma_{K,3D-ph}T^3$ 
is the Kapitza boundary conductance at temperature T.

These highly nonlinear equations were solved approximately by numerical methods, 
yielding the position dependent temperature 
profiles shown in Figs. ~\ref{fig6}(b) and (c), respectively 
for S2 and S4.  The fitting to $I_r$ versus temperature 
yielded the curves in Figs. ~\ref{fig3}(b) and (c), with $L_o$ fixed at the value from wire S1, 
of $3.65 \times 10^{-8} W\Omega/K^2$.  The critical value $x_c$ was 
found to shiftly slightly toward the center, to $0.23 L$  ($0.46(L/2$) rather than $L/4$ ($0.5(L/2)$).  
The best fit is for $\gamma = 5$, corresponding to 
the electron-3D phonon energy relaxation as the limiting step.  Fits of slightly lower quality 
can be achieved for $\gamma = 4$ or $3$.  On the other hand, the numerical values for the 
coefficient $\beta$ yielded values for the parameters, which point to Kapitza boundary 
resistance at the Al nanowire-InP interface as the limiting path to phonon conduction.  

For $\gamma = 5 $ the extracted electron-3D-phonon relaxation rate $\Sigma_{e-3D-ph}$ 
for S2 is $\sim 6 \times$ 
as large as the established value $\sim 2 \times 10^9W/m^3K^5$\cite{Hysteresis.SNS} and 
is $\sim 11 \times$ for S4.  The values are thus inconsistent.  
We are forced to consider the possibility that the Al phonons are reduced in dimensions down 
to 1D.  An enhancement of $\sim \Theta_D/T (a/w) \sim 9$ can be expected per dimension reduced, 
where $a$ is the lattice constant, yielding a factor $\sim 81$, far larger than the measured 
enhancement!  Instead, within this scenario, one expects the limiting step to be the Kapitza 
boundary resistance between the Al nanowire and the InP ridge.
 
For Kapitza boundary limited conduction, one may expect 
$\gamma = 3$ rather than $\gamma =4$ due to reduced-dimension 2D InP phonons,
despite the poorer quality fit.  An  
enhancement factor of $\sim 9$ over the known 3D value should be present from 
the reduction in dimension by 1.  As a reference, we 
use $\sigma_{K,3D-ph} \sim 20 W/m^2K^4$, obtained for Au on GaAs rather than 
Al on InP \cite{MohantyRoukes}.  Forcing $\gamma = 4$ yield the parameter 
value $\sigma_{K,3D-ph} \sim 190 W/m^2K^4$, which is 10 times the reference 
value.  Instead, after conversion of the reference value to account for 2D phonons, 
the enhanced value 
of $\sigma_{K,2D-ph} \sim 280 W/m^2K^3$ is consistent with the fitted values of $300 W/m^2K^3$ 
and $330 W/m^2K^3$, respectively, for S2 and S4 (with $\gamma = 3$).  To make certain this 
picture of Kapitza boundary limited conduction is consistent, we need to 
ensure the ridge phonon thermal conduction is larger.  An estimate of the phonon thermal 
conductivity of the InP ridge itself yields a 
lower bound of $700 W/m^2K^3$, corresponding to the case of a 
very short, ridge-width limited phonon mean free path $\sim 8 nm$.  
This rules out phonon conduction in the InP ridge as the limiting step, as 
required.  Finally, as a reference, we estimate the conductivity in the 
{\it absence} of any electronic contribution.  This yield a value {\it roughly 
double} the above values for the 2D phonon Kaptiza boundary resistance 
coefficient.

It is worthwhile to re-emphasize the evidence for reduced dimension Kaptiza 
boundary conduction as the limiting step, based on a direct comparison of the 
fitting parameter values for S2 and S4.  Whereas, 
$\Sigma_{e-3D-ph}$ for electron-phonon relaxation 
limited heat removal shows a discrepancy between S2 and S4 of a factor $1.2/2.2 \sim 
1.8$, the values for the reduced dimension $\sigma_{K,2D-ph}$ for 2D InP 
ridge phonon, Kapitza boundary limited thermal conduction are within $10 \%$ of 
each other!  This, in conjunction with the discrepancy with the known reference value for 
$\Sigma_{e-3D-ph}$, helps establish the Kapitza boundary resistance limited scenario.

Based on the detailed analysis presented in this work, we establish that in the 
retrapping process, the longer wires S2 and S4 require phonons to 
contribute to heat removal, in addition to the electronic thermal conduction, 
while for the short S1, electronic conduction alone is sufficient.  
The reasonable fits using sensible parameters demonstrate that 
it is possible to achieve an understanding of the heating-induced hysteresis 
for nanowires S1, S2 and S4, based on heating within the normal regions, 
while at the same time account for the observed differences.



\begin{references}



\bibitem{Giordano.PRL}
N. Giordano, 
Phys. Rev. Lett. {\bf 61}, 2137 (1988).

\bibitem{Nature}
A. Bezryadin, C.N. Lau, and M. Tinkham, 
Nature {\bf 404}, 971 (2000).

\bibitem{Lau.PRL}
C.N. Lau, N. Markovic, M. Bockrath, A. Bezryadin, and M. Tinkham, 
Phys. Rev. Lett. {\bf 87}, 217003 (2001).

\bibitem{Highfield.B}
A. Rogachev, A.T. Bollinger, and A. Bezryadin, 
Phys. Rev. Lett. {\bf 94}, 017004 (2005).

\bibitem{Fabio.PRL}
F. Altomare, A.M. Chang, M.R. Melloch, Y, Hong, and C.W. Tu, 
Phys. Rev. Lett. {\bf 97}, 017001 2006).


\bibitem{HysteresisIV}
M. Tinkham, J.U. Free, C.N. Lau, and N. Markovic, 
Phys. Rev. B {\bf 68}, 134515 (2003).


\bibitem{NaturePhysics}
M. Sahu, M.H. Bae, A. Rogachev, D. Pekker, 
T.C. Wei, N. Shah, P.M. Goldbart, A. and Bezryadin, 
Nat. Phys. {\bf 5}, 503 (2009).

\bibitem{PengPRL}
P. Li, Phillip M. Wu, Yuriy Bomze, I. Borzenets, G. Finkelstein, and A.M. Chang, 
arXiv: 1006.420.
\bibitem{Goldbart}
N. Shah, D. Pekker, and P.M. Goldbart, 
Phys. Rev. Lett. {\bf 101}, 207001, (2008).

\bibitem{Little}
W.A. Little, 
Phys. Rev. {\bf 156}, 396 (1967).

\bibitem{LA}
J.S. Langer and V. Ambegaokar, 
Phys. Rev. {\bf 164}, 498 (1967).

\bibitem{MH}
D.E. McCumber, and B.I. Halperin, 
Phys. Rev. B \{bf 1), 1054 (1970).

\bibitem{QPStheory.Zaikin1}
A.D. Zaikin, D.S. Golubev, A. van Otterlo, and G.T. Zim\'anyi, 
Phys. Rev. Lett. {\bf 78}, 1552 (1997).

\bibitem{QPStheory.Zaikin2}
D.S. Golubev, and A.D. Zaikin, 
Phys. Rev. B {\bf 64}, 014504 (2001).

\bibitem{sergei}
S. Khlebnikov, and L.P. Pryadko, 
Phys. Rev. Lett. {\bf 95}, 107007 (2005).

\bibitem{CurrentJJ}
J.E. Mooij, and Y.V. Nazarov,
Nat. Phys. {\bf 2}, 169 (2006).

\bibitem{nanowire.qbit}
J.E. Mooij and C.J.P.M. Harmans,
New Journal of Physics {\bf 7}, 219 (2005).

\bibitem{Hysteresis.SNS}
H. Courtois, M. Meschke, J.T. Peltonen, and J.P. Pekola, 
Phys. Rev. Lett. {\bf 101}, 067002 (2008).





\bibitem{TinkhamBook}
M. Tinkham, {\it Introduction to superconductivity}, International series 
in pure and applied physics (McGraw Hill, New York, 1996), 2nd ed.

\bibitem{Fabio.APL}
F. Altomare A.M. Chang, M.R. Melloch, Y. Hong, and C.W. Tu, 
Appl. Phys. Lett. {\bf 86}, 172501 (2005).

\bibitem{MQT.Clarke}
M.H. Devoret, J.M. Martinis, and J. Clarke, 
Phys. Rev. Lett. {\bf 55}, 1908 (1985).


\bibitem{Collapse1}
V.M. Krasnov, T. Bauch, S. Intiso, E. H\"urfeld, 
T. Akazaki, H. Takayanagi, and P. Delsing, 
Phys. Rev. Lett. {\bf 95}, 157002 (2005).

\bibitem{Collapse2}
J.M. Kivioja, T.E. Nieminen, J. Claudon, O. Buisson, 
F.W.J. Hekking, and J.P. Pekola, 
Phys. Rev. Lett. {\bf 94}, 247002 (2005).





\bibitem{Klapwijk.PRB}
G. R. Boogaard, A. H. Verbruggen, W. Belzig, and T. M. Klapwijk, Phys. Rev. B {\bf 69}, 
220503 (2004).

\bibitem{MohantyRoukes}
P. Mohanty, D. A. Harrington, K. L. Ekinci, Y. T. Yang, 
M. J. Murphy, and M. L. Roukes, Phys. Rev. B {\bf 66}, 085416 (2002).


\end{references}
\end{document}